\PassOptionsToPackage{hidelinks}{hyperref}
\documentclass[useAMS, usenatbib]{mnras}
\usepackage{epsfig}
\usepackage{color} 
\usepackage{natbib}
\usepackage{deluxetable}
\usepackage{amssymb}
\usepackage{rotating}
\usepackage{graphicx}
\usepackage{hyperref}
\usepackage{graphicx}
\usepackage{multicol}
\usepackage{blindtext}
\usepackage{comment}
\makeatletter
\def\footnoterule{\kern-3\p@
  \hrule \@width 2in \kern 2.6\p@} 
\makeatother

\def\apjl{{ApJ}}%
\def\apjs{{ApJS}}%
\def\aap{{A\&A}}%
\title[Evidence of underdeveloped torus $\&$ BLR in WLQ] 
{Evidence of under-developed torus and broad-line region of weak emission line quasars based on their spectral energy distribution} 
\author[Ritish Kumar, Hum Chand, Ravi Joshi] {Ritish Kumar$^{1}$\thanks{E-mail: ritishshield@gmail.com}, Hum Chand$^{1}$, Ravi Joshi$^{2}$\\$^{1}$ Department of Physics and Astronomical Science,  Central University of Himachal Pradesh.\\
    $^{2}$Indian Institute of Astrophysics, Koramangla, Bangalore, 560034, India\\
}
\begin{document}
\date{Accepted ---. Received ---; in original form ---}

\pagerange{\pageref{firstpage}--\pageref{lastpage}} \pubyear{2022}

\maketitle

\label{firstpage}
\begin{abstract}
To unravel the dominant cause of the weak emission line in a subset of optically selected radio-quiet `weak emission line quasars' (WLQs), we have investigated the possibility of an underdeveloped broad line region (BLR). For this, we have modeled spectral energy distributions (SED) of 61 WLQs by using their optical and infrared (IR) photometric observations from SDSS and WISE respectively. SED fit consists of various emission components, including the luminosity from the dusty torus ($L_{tor}$). For comparison with the normal quasar, we have used a control sample of  55 QSOs for each WLQs matching in emission redshift and SDSS r-band. Based on our measurement of $L_{tor}$, we found a decrement of $42\pm2$\% in IR-luminosity in WLQs w.r.t the control sample of normal QSOs. 
Using $L_{tor}$/$L_{bol}$ as the measure of torus covering factor ($CF_{tor}$) we found a similar decrement in WLQs covering factor, with their $CF_{tor}$ distribution being  significantly different w.r.t. the normal QSOs with a KS-test $P_{null}$ of $4.27 \times 10^{-14}$. As dusty torus and BLR covering factors are expected to be of a similar order in AGN, our results suggest that the BLR in the WLQs is underdeveloped and could be a dominant cause of the weakness of their emission line. As a result, our analysis gives support to the models of WLQs based on the evolution scenario being in  an early stage of AGNs. 
\end{abstract}
\section{Introduction} 
\label{Sec 1}
The key ingredients of a unified model of the AGN central engine are composed of a supermassive black hole in the center along with the various substructures including the accretion disk, broad line region (BLR), and dusty torus. The dusty torus plays a crucial role in absorbing radiation from the central part and re-emitting it in IR, as well as explaining the distinct characteristics of Type-1 and Type-2 AGN based on their different orientation relative to the line of sight \citep[][]{antonucci1993unified, Urry1995PASP..107..803U}. 
The BLR region has high-velocity gas (FWHM = 1000–20,000 kms$^{-1}$) being embedded in the gravitational potential well of a quasar’s supermassive black hole (SMBH). The high energy photons (UV and/or X-ray) from the accretion disk and corona photo-ionize the gas in BLR, giving rise to prominent broad emission lines with a rest-frame equivalent width (EW$_r$) of about 50–110 \AA$~$for typical Ly$\alpha+$N~{\sc v} emission line in quasars \cite[e.g., see][]{Schneider1991AJ....101.2004S,francis1993ultraviolet,osmer1994luminosity,warren1994wide,zheng1997composite,brotherton2001composite,dietrich2002continuum}. However, these lines can either disappear or appear much weaker (typically, EW$_r < 15$ \AA$~$for Ly$\alpha+$ N~{\sc v}) for a subclass of BL Lac objects (BLOs) in which optical/UV emission is dominated by the doppler boosted non-thermal continuum from the relativistic jet \citep[e.g., see][]{begelman1984theory}. 
As BLOs are jet dominated \cite[e.g., see][]{Urry1995PASP..107..803U}, they are radio loud having radio to optical flux density ratio, R\footnote{Radio loudness (R) is usually parameterized by the ratio of flux densities at 5 GHz 
and at 2500~\AA~in the rest frame, being R $>$ 10 and R $<$ 10 for radio-loud and radio-quiet respectively \citep[e.g., see][]{Kellermann1989AJ.....98.1195K}.}$>$10. 
However, with the advent of large spectroscopic surveys, such as the Sloan Digital Sky Survey \citep[SDSS; e.g.,][]{York2000AJ....120.1579Y} and the Two$-$Degree Field QSO Redshift Survey \citep[2QZ;][]{boyle20002df}, a peculiar or new class of hundreds of high-redshift (mostly at $z>2$) radio-quiet quasars (RQQs) displaying  exceptionally weak or even, in some cases completely missing, emission lines were discovered \cite[e.g., see][]{Collinge2005AJ....129.2542C,Fan2006AJ....131.1203F,Anderson2007AJ....133..313A,Plotkin2010AJ....139..390P,Plotkin2010ApJ...721..562P,Wu2011ApJ...736...28W,Meusinger2014A&A...568A.114M}. These objects are commonly known as weak emission line quasars (WLQs), various models have been proposed for the observed weak emission lines, including 
(i) radiatively inefficient accretion flow (RIAF) leading to intrinsically weaker optical/UV continuum radiation \citep[][]{yuan2004nature}, 
(ii) soft ionising continuum causing the weakness of emission lines in BLR \citep[][]{Leighly2007ApJS..173....1L,Leighly2007ApJ...663..103L,Laor2011MNRAS.417..681L,Wu2011ApJ...736...28W,luo2015x}, and (iii) anemic or unusual BLR \cite[][]{shemmer2010weak,Nikolajuk2012MNRAS.420.2518N}, by postulating WLQs being in the early evolutionary phase of quasars \cite[e.g., see][]{Hryniewicz2010MNRAS.404.2028H, Liu2011ApJ...728L..44L}.\par   
On the observation side, many programs have also been carried out to either confirm or refute many of the models mentioned above. For instance, the observed high luminosity of WLQs \citep[e.g., see][]{Meusinger2014A&A...568A.114M} excludes the possibility of radiatively inefficient accretion flow as the leading cause behind the origin of their weak emission lines. 
For models based on soft ionizing photons, the possible mechanisms are (i) a cold accretion disk around a supermassive black hole with $M_{BH}~> 3.6\times 10^9 M_\odot$ for non-rotating and $> 1.4\times 10^{10}M_\odot$ for maximally rotating black hole \citep{Laor2011MNRAS.417..681L}, though based on the measured SMBH mass of WLQs, it also appears to be an unlikely scenario \citep[e.g., see][]{Meusinger2014A&A...568A.114M, Plotkin2015ApJ...805..123P}, (ii) extremely high accretion rate inducing a soft ionizing continuum \citep{Leighly2007ApJS..173....1L, Leighly2007ApJ...663..103L}, however, the observed low fraction of WLQs with such a required high accretion rate also excludes this possibility \citep[e.g., see][]{Nikolajuk2012MNRAS.420.2518N}. Another  possibility is based on the high Eddington ratio in a geometrically thick accretion disk which creates a shielding gas that prevents the BLR from getting photo-ionized by a central continuum source \citep[e.g.,see ][]{Wu2011ApJ...736...28W,luo2015x,ni2018connecting,paul2022connecting}. Sources with weak emission lines may also be obscured AGNs \citep{1988ApJ...328..569U}, where spectropolarimetry is found useful to confirm or refute such possibility of obscuration of the continuum/BLR emission from the direct view \citep[e.g., see][]{1988ApJ...331..332G}.\par
Alternatively, the BL-Lac nature of WLQs is also tested by comparing the intranight optical variability duty cycle \cite[e.g., see ][]{Gopal2013MNRAS.430.1302G, Chand2014MNRAS.441..726C, Kumar2015MNRAS.448.1463K,kumar2016intranight,kumar2017multi} and nature of their optical polarization \cite[e.g., see][]{Smith2007ApJ...663..118S,DiamondStanic2009ApJ...699..782D, Heidt2011A&A...529A.162H,2018MNRAS.479.5075K}. Unlike the  blazars with higher INOV duty cycle ($\sim$ 30 - 50\%) \cite[e.g., see][]{2018BSRSL..87..281G} and higher polarization value, the WLQs are found to have a INOV duty cycle of $\sim$ 5\% and polarization of $< 3\%$ , thus are more like a normal radio-quite quasars. 
These observational pieces of evidence seems to favour the scenario where WLQs might belong to the early phase of AGNs' lifetime rather than being the radio-quiet counterparts of BL Lac. 
In this evolutionary scenario, the radiation from the spherical cocoon of gas can easily escape through the axis of angular momentum and hence help in the formation of doughnut-shape structure, finally leading to the dusty torus \cite[e.g., see ][their figure 5]{Liu2011ApJ...728L..44L}. Accordingly, the accretion disk in WLQs will be relatively recently established, and hence the BLR is unlikely to be significantly developed yet \citep[e.g.,  see][]{Hryniewicz2010MNRAS.404.2028H, Liu2011ApJ...728L..44L, andika2020probing}. 
This will have its consequences on the covering factor of the BLR in WLQs, to be at least an order of magnitude smaller compared to the normal QSOs, as found in the study of \citet[][]{Nikolajuk2012MNRAS.420.2518N} based on the ratios of high-ionization line and low-ionization line regions. 
Additionally, \cite{gaskell2007ngc,gaskell2009broad} has argued that the covering factor of BLR and dusty torus has to be the same. This is due to the fact that lower covering factor of torus will lead to BLR in absorption which is not supported by observation \citep[e.g., see  ][]{1989ApJ...342...64A,1992ApJ...400..435K}. 
Similarly, the torus with a covering factor higher than BLR will be unable to exist due to direct radiation from the central source on its portion not shielded by the BLR \cite[e.g., see][]{gaskell2009broad}. For instance, \citet{1993ApJ...404L..51N} proposed that the outer boundary of the BLR is set by dust formation which is also confirmed by IR reverberation mapping \citep[e.g., see][]{2006ApJ...639...46S,gaskell2007ngc}, predicting that the covering factor of BLR and dusty torus should be similar. Therefore, in the case of WLQs, if they have an underdeveloped BLR, then the smaller covering factor of the dusty torus will have additional observational consequences in the infrared (IR) band, viz., the reduction of its IR emissivity in comparison to the normal QSOs. 
To test this hypothesis, a comparison of the infrared spectral energy distribution (SED) of WLQs and normal QSOs matched in their optical luminosity and redshift will be very useful. In this context, \citet{DiamondStanic2009ApJ...699..782D} has reported a reduction of about 30–40\% in the IR luminosity of two WLQs (viz., SDSS J140850.91+020522.7 with EW(C IV) = 1.95~\AA~and SDSS J144231.72+011055.2 with EW(C IV) = 16.9~\AA). Similarly, \citet{zhang2016covering} have also compared the IR luminosity of normal QSOs with WLQs by using SED fitting, and they have found that WLQs and normal QSOs are statistically similar, though their results are consistent with the evolution scenario. Also, their model of SED fitting was very simplistic, consisting of only the best-fit model of power law and a single-temperature black body. In such a simplistic model, any difference in the IR-luminosity originating from the dusty torus might get diluted in the absence of a proper decomposition of the various emission components of AGN nuclei, such as the AGN inner accretion disk, dusty torus, host galaxy, and the cold dust in star-forming regions. Therefore, to confirm or refute the scenario of the less developed BLR as the cause of the weak emission line in WLQs based on the observed IR emission from the dusty torus, it becomes important to carry out the SED fitting of the large sample of the WLQs by properly decomposing the various components of emissions. This forms the main motivation of the present work. 
Here we used IR observations of WLQs from the Wide-field Infrared Survey Explorer \citep[WISE;][]{wright2010wide} band in conjunction with their SDSS observation in the optical band, to model the SED of each WLQ and compare it with the SED of control sample of normal QSOs matched in redshift and SDSS r-band magnitude.\par 
The paper is organized as follows: Section~\ref{section:sample} describes our sample selection of WLQs and control sample of normal QSOs. In Section~\ref{section:analysis}, we present analysis and results based on SED fitting of our sample, followed by a discussion and conclusions in Section~\ref{section:discussion}. 
In this paper we assume that $H_0 = 70 kms^{-1}Mpc^{-1}$ , $\Omega_{\lambda} = 0.7$, and $\Omega_M = 0.3$.
\section{Sample of WLQs}
\label{section:sample}
Our parent sample consists of WLQs selected from two catalogs based on the SDSS Data Release 7 \citep[DR-7][]{Abazajian2009ApJS..182..543A} given by \citet[][hereafter PL10]{Plotkin2010AJ....139..390P} and \citet[][hereafter MB14]{Meusinger2014A&A...568A.114M}. In table 6 of the PL10 catalog, they have given a list of 86 high-confidence WLQs based on featureless optical spectra and radio-quietness (i.e. $R< 10$). In the catalog of MB14, they employed machine learning data mining techniques to the large database of quasars in the SDSS DR7 pipeline \citep[DR-7;][]{Abazajian2009ApJS..182..543A}. This is followed by manual inspection as well as imposing rest-frame equivalent-width thresholds: EW(Mg~{\sc ii}) $<11$~\AA~and EW(C~{\sc iv}) $<4.8$~\AA, leading to a well-defined sample of 46 WLQs.
Out of them, 9 were found to be common with the 86 WLQs taken from PL10, leading to the addition of only 37 WLQs from the MB14 sample. All the 86 WLQs in the PL10 are radio-quiet, having a radio-loudness parameter, R $<$ 10. However, among the 37 sources considered from the MB14 catalog, 15 sources were excluded due to their radio-loudness parameter R$>$10. 
As a result, we are left with 108 WLQs (86 from PL10 and 22 from MB14) as our initial parental sample. We further checked for the availability of the WISE data for our sample of these 108 WLQs in the compilation of \citet{paris2018sloan} in the SDSS DR14 quasar catalog and found that the WISE data is available for 98 WLQs. We have also carried out a visual inspection of the SDSS spectrum of these 98 WLQs and noted that 5 sources  were genuinely identified as galaxies and hence reduced our sample to 93 sources. 
As an extra check for the genuine extra-galactic nature of our sources, we also checked the proper motion of our sources based on the proper motion catalog of \citet{Monet2003AJ....125..984M}. For this, a criterion of proper motion to be either zero or consistent with zero at $<$ $2.5\sigma$ level is adopted \citep[e.g., see][]{Kumar2015MNRAS.448.1463K}. 
This criterion excluded 12 sources with significant proper motion and led us to a sample of 81 sources. Further, we have also excluded 4 sources belonging to crowded fields due to the high possibility of contamination of their photometric fluxes (to be used in our SED fit), leading to a sample of 77 sources. For the comparison of the SED of the WLQs with the normal quasars, we have made the control sample of normal quasars matching in redshift and r-band magnitude with WLQs within the tolerance of $ \le 0.01$ and $ \le 0.6$ mag, respectively. The chosen tolerance limits are found optimal to get  55 normal QSOs control samples for each WLQ except 8 WLQs. These 8 WLQs belong either to very low (0.04$<$z$<$0.06) or very high (3.5$<$z$<$6.5) redshift ranges, resulting in  scarcity of sources for the control sample within the tolerance limits. 
This led to our sample of 69 WLQs sources along with a control sample of 55 normal QSOs for each of them, with details such as name, RA, DEC, optical (u, g, r, i, z) and infrared (W1, W2, W3, W4) magnitude, as given in Table~ \ref{tab:Table_1} for WLQs as well as in Table~\ref{tab:Table_2} for the median properties of the control sample of normal QSOs corresponding to each WLQs. 
\begin{table*}
\begin{centering}
\caption{ Basic parameters of the 69 WLQs in our sample.} 
\label{tab:Table_1}
\renewcommand{\arraystretch}{1.5}
\renewcommand{\tabcolsep}{1.5mm}
{\tiny
\begin{tabular}{rlrrrrrrrrrrrr}
\hline
 \multicolumn{1}{c}{SN.}  & \multicolumn{1}{c}{Source Name}  &
 \multicolumn{1}{c}{R.A}  & \multicolumn{1}{c}{Dec} &
 \multicolumn{1}{c}{z} & \multicolumn{1}{c}{$u_{mag}$$\pm$err} &
 \multicolumn{1}{c}{$g_{mag}$$\pm$err} &  \multicolumn{1}{c}{$r_{mag}$$\pm$err} &
 \multicolumn{1}{c}{$i_{mag}$$\pm$err} &  \multicolumn{1}{c}{$z_{mag}$$\pm$err} &
 \multicolumn{1}{c}{$W_{1}$$\pm$err} &  \multicolumn{1}{c}{$W_{2}$$\pm$err} &
 \multicolumn{1}{c}{$W_{3}$$\pm$err} &  \multicolumn{1}{c}{$W_{4}$$\pm$err} 
 \\
 \hline
 1 & J001444.03$-$000018.5 &   3.68 &  $-$0.005 & 1.549  & 18.2 $\pm$0.02 & 17.99$\pm$0.03 & 17.87$\pm$0.03 & 17.72$\pm$0.02 &   17.58$\pm$0.02 &   14.98$\pm$0.04 &   13.96$\pm$0.04 &   11.29$\pm$0.2  &    8.34$\pm$0.04 \\
 2 & J001514.88$-$103043.6 &   3.81 & $-$10.51  & 1.170  & 19.59$\pm$0.05 & 19.57$\pm$0.03 & 19.23$\pm$0.02 & 19.1$\pm$0.02 &   19.03$\pm$0.06 &   15.71$\pm$0.05 &   14.79$\pm$0.07 &   12.07$\pm$0.38 &    8.81$\pm$0.05 \\
 3 & J001741.87$-$105613.2 &   4.42 & $-$10.94  & 1.806  & 19.22$\pm$0.03 & 19.00   $\pm$0.03 & 18.8 $\pm$0.02 & 18.64$\pm$0.02 &   18.63$\pm$0.04 &   16.06$\pm$0.06 &   14.73$\pm$0.06 &   11.51$\pm$0.21 &    8.75$\pm$0.06 \\
--- & \multicolumn{1}{c}{---} &\multicolumn{1}{c}{---} &\multicolumn{1}{c}{---} &\multicolumn{1}{c}{---} & \multicolumn{1}{c}{---} &\multicolumn{1}{c}{---} &\multicolumn{1}{c}{---} &\multicolumn{1}{c}{---} &\multicolumn{1}{c}{---} &\multicolumn{1}{c}{---} &\multicolumn{1}{c}{---} &\multicolumn{1}{c}{---} &\multicolumn{1}{c}{---}  \\
\hline
\multicolumn{14}{{|p{1.8\columnwidth}|}}{\textbf{Note:} The entire table is available in online version. Only a portion of this table is shown here to display its form and content.}
\end{tabular}
} 
\end{centering}
\end{table*}  
\begin{table*}
  \centering
      {
        \setlength{\tabcolsep}{1.2pt}
\caption{ Median values of  the basic parameters of the  control sample corresponding to the 69 WLQs in our sample.}
\label{tab:Table_2}
\renewcommand{\arraystretch}{1.5}
\renewcommand{\tabcolsep}{2mm}
{\tiny
\begin{tabular}{rlrrrrrrrrr}
\hline
 \multicolumn{1}{c}{SN.}  & \multicolumn{1}{c}{Source Name*}  &
 \multicolumn{1}{c}{$u_{mag}\pm$err} &
 \multicolumn{1}{c}{$g_{mag}\pm$err} &  \multicolumn{1}{c}{$r_{mag}\pm$err} &
 \multicolumn{1}{c}{$i_{mag}\pm$err} &  \multicolumn{1}{c}{$z_{mag}\pm$err} &
 \multicolumn{1}{c}{$W_{1}\pm$err} & \multicolumn{1}{c}{$W_{2}\pm$err} &
 \multicolumn{1}{c}{$W_{3}\pm$err} & \multicolumn{1}{c}{$W_{4}\pm$err}
 \\
 \hline
 1 & J001444.03$-$000018.5 &18.40 $\pm$0.00 &18.09$\pm$0.00  &17.96$\pm$0.00  &17.74$\pm$0.00  &17.74$\pm$0.00  &14.55 $\pm$0.13 & 13.15 $\pm$0.13 &10.06 $\pm$0.13 &7.87 $\pm$0.13 \\
2 & J001514.88$-$103043.6 &19.70  $\pm$0.01 &19.55$\pm$0.00  &19.22$\pm$0.00  &19.2 $\pm$0.00  &19.27 $\pm$0.01 &15.53 $\pm$0.13 &14.33 $\pm$0.13 &11.48 $\pm$0.13 &8.66 $\pm$0.13 \\
3 & J001741.87$-$105613.2 &19.06$\pm$0.00  &18.89$\pm$0.00  &18.80 $\pm$0.00  &18.52$\pm$0.00  &18.51$\pm$0.00  &15.48 $\pm$0.13 & 14.20 $\pm$0.13 &10.94 $\pm$0.13 &8.43 $\pm$0.13 \\
--- & \multicolumn{1}{c}{---} &\multicolumn{1}{c}{---} &\multicolumn{1}{c}{---} &\multicolumn{1}{c}{---} & \multicolumn{1}{c}{---} &\multicolumn{1}{c}{---} &\multicolumn{1}{c}{---} &\multicolumn{1}{c}{---} &\multicolumn{1}{c}{---} &\multicolumn{1}{c}{---}  \\
\hline
\multicolumn{11}{l}{\textbf{Note:}The entire table is available in online version. Only a portion of this table is shown here to display its form and content.}\\
\end{tabular}
}  
}
\end{table*}
\begin{figure}
   {\includegraphics[width = 9.3 cm]{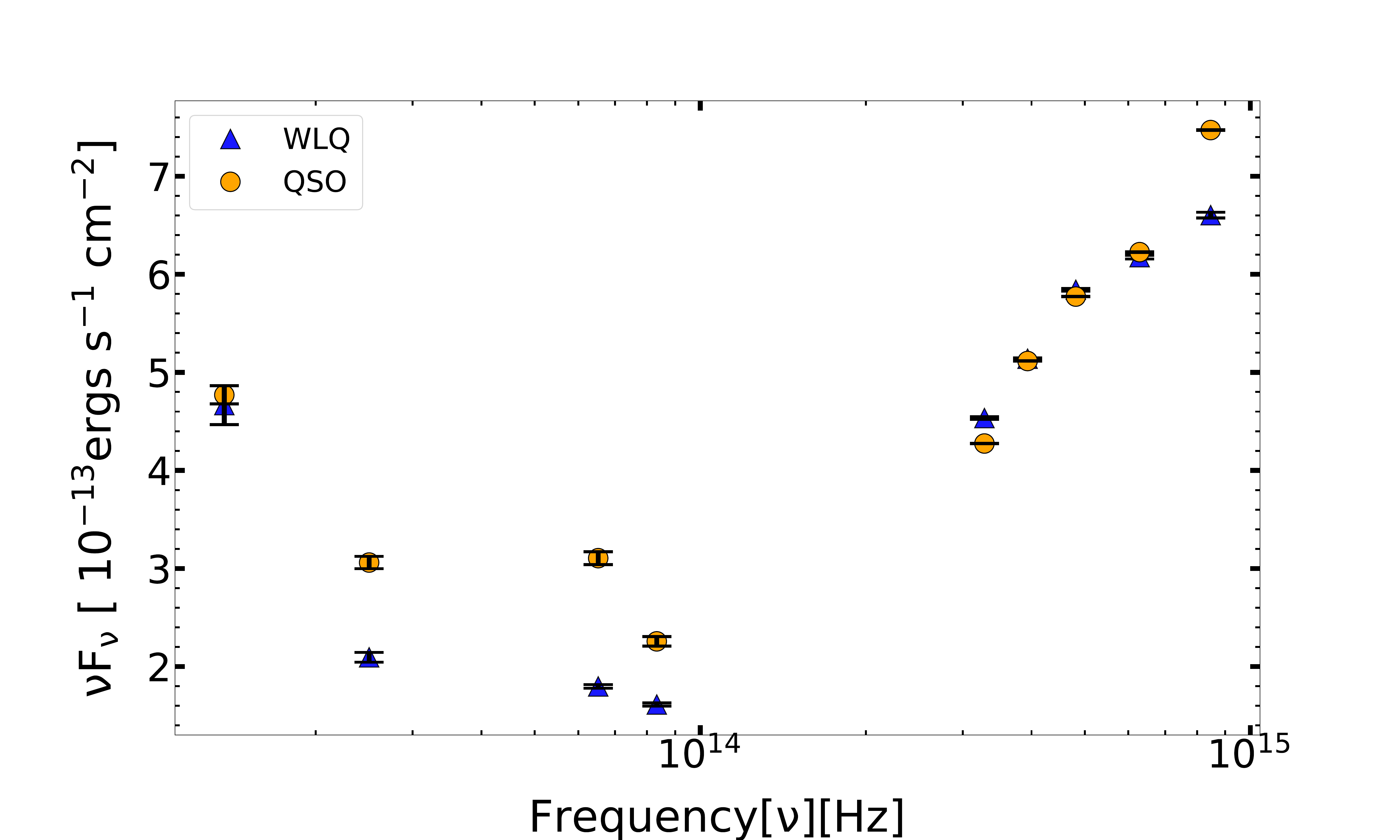}}
   \caption{Comparison of median values of the observed flux ({$\nu$}$F_{\nu}$) of the WLQs sample and the corresponding median value of the control sample of normal quasars. The triangle denotes the WLQs whereas the solid circle denote the normal quasars. The error bars in the median values, obtained by proper error propagation, are smaller than the symbol size.} 
   \label{fig:fig1}
\end{figure}
\begin{figure}
   \centering
   {\includegraphics[width = 8.8cm]{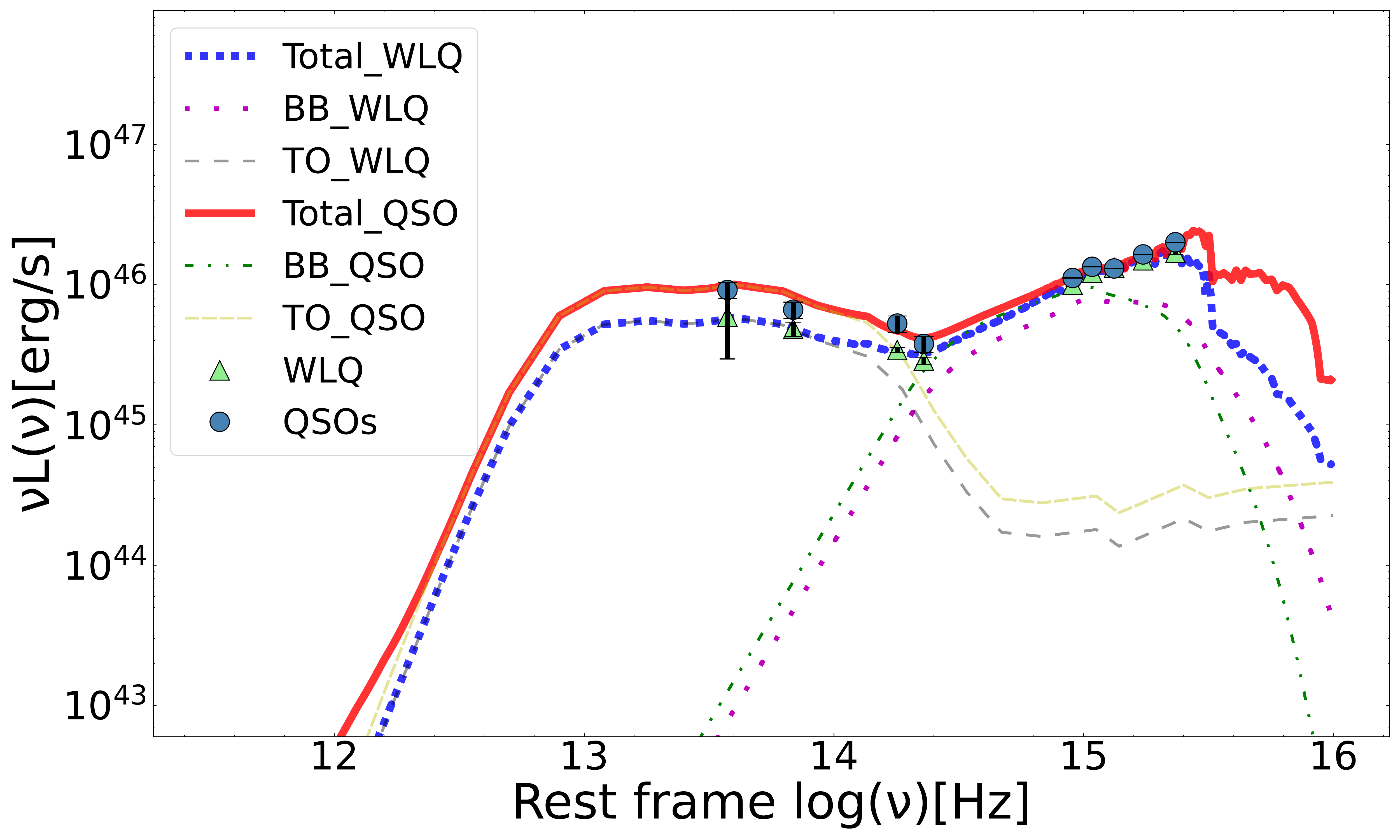}}
   \caption{Illustration of the best fit SED comparison of different  components of emission in one of the WLQ from our sample, namely J141141.96$+$140233.9 (solid triangle, thick doted line) along with SED fit to its corresponding median flux value of the control sample of normal QSOs (solid circle, thick line).}
   \label{fig:sed}
\end{figure}
\begin{figure}
   {\includegraphics[width = 8.6 cm]{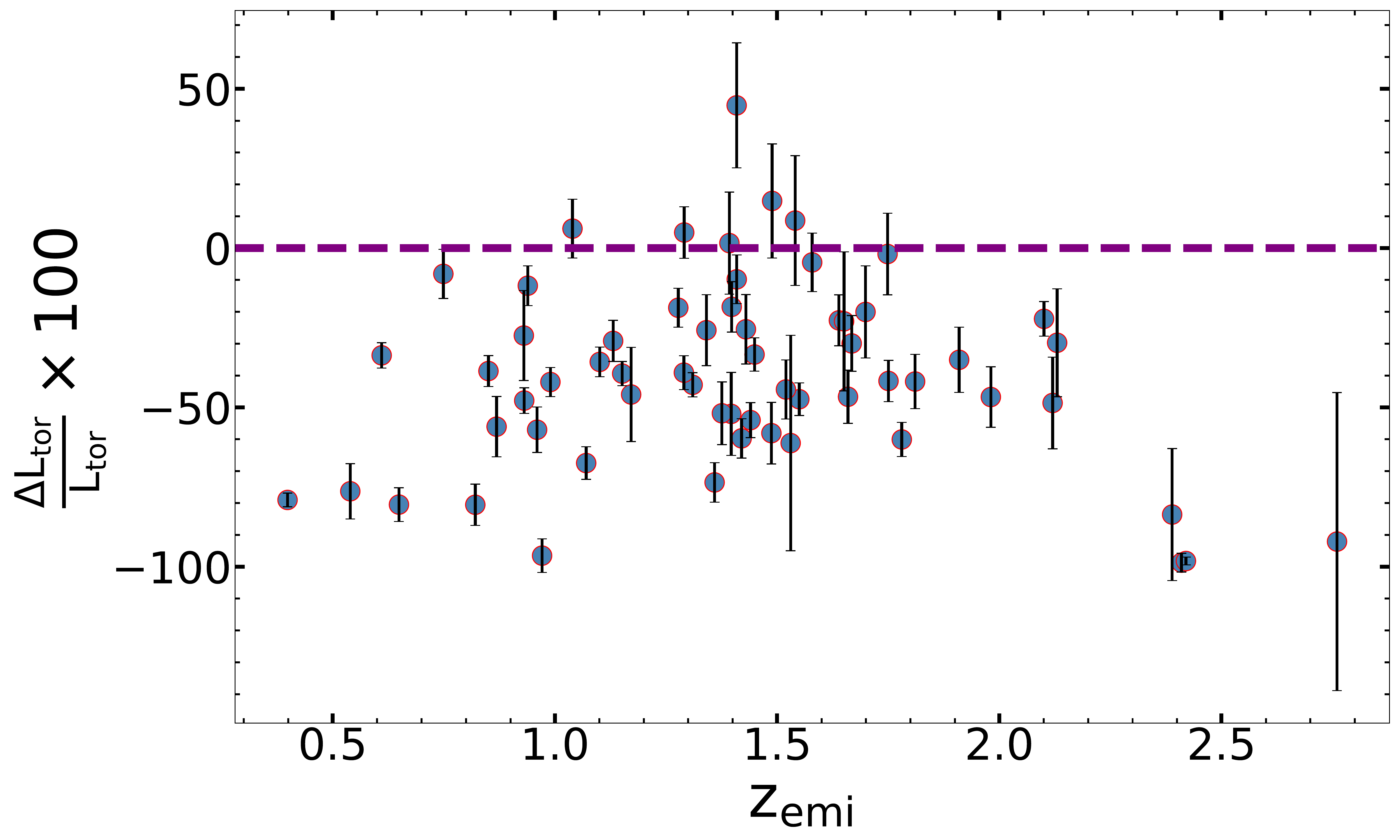}}
   \caption{The plot show $\Delta L_{tor}/L_{tor}(\equiv  [L_W^{tor}-L_Q^{tor}]/L_{Q}^{tor})$ verses emission redshift ($z_{emi}$), computed for each member of the WLQ with respect to its control sample of normal QSOs. It can be noticed that (i) except few outliers, all the WLQs shows smaller torus luminosity in comparison to their redshift and r-band magnitude matched sample of normal QSOs, (ii) there is no significant trend in the $\Delta L_{tor}/L_{tor}$ as a function of emission redshift. }
   \label{fig:3}
\end{figure}
\begin{figure}
   {\includegraphics[width = 8.7 cm]{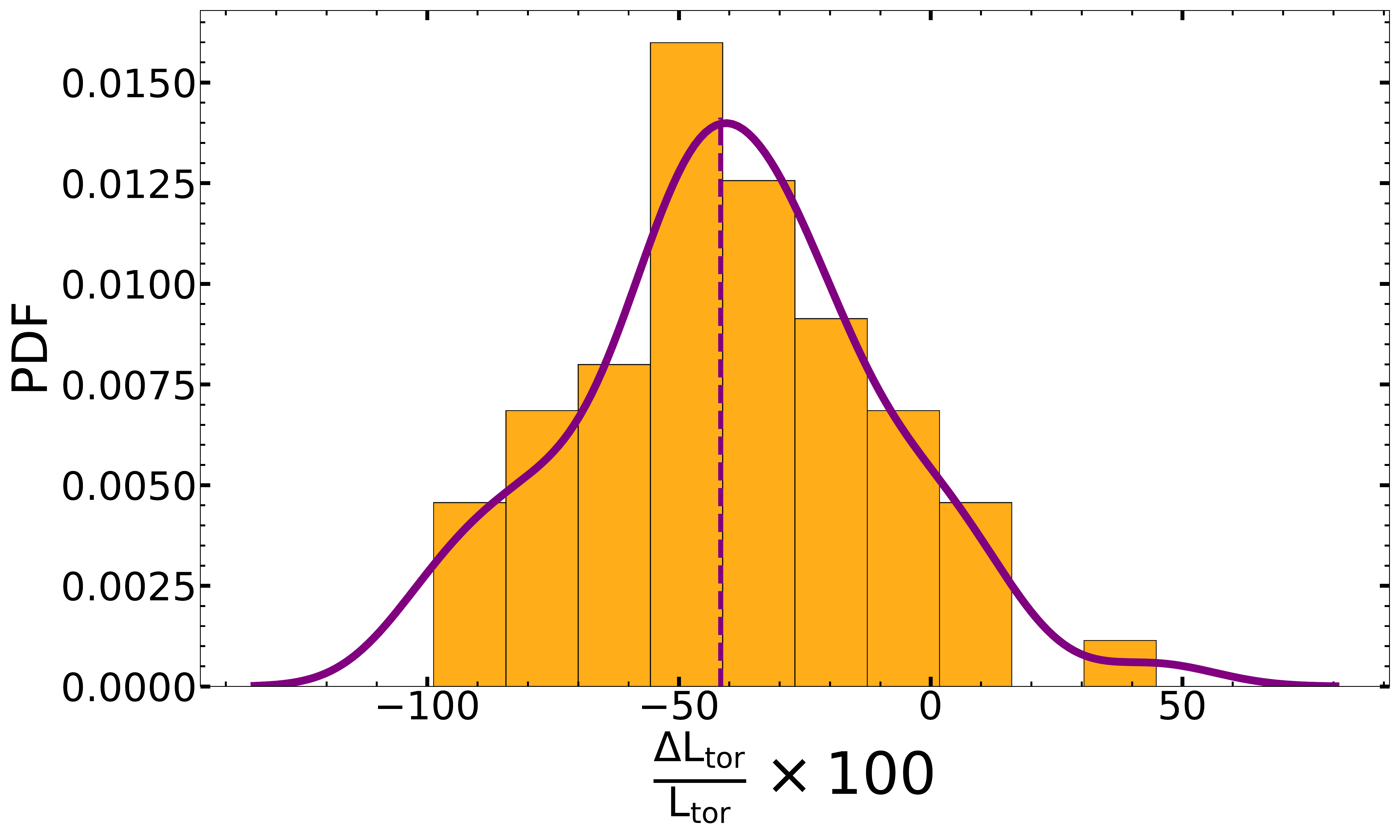}}
   \caption{The kernel smooth probability distribution function (PDF) of the percentage deviation of torus luminosity of WLQs with respect to the control sample of QSOs ($\Delta L_{tor}/L_{tor}$). The median percentage deviation $\Delta L_{tor}/L_{tor}$ is found to be $-42\pm 2 \%$. The Gaussian function fit to this  distribution results in a 
   typical 1$\sigma$ uncertainty of 29\%.
   }
   \label{fig:5}
\end{figure}
\begin{figure}
   {\includegraphics[width = 8.8 cm]{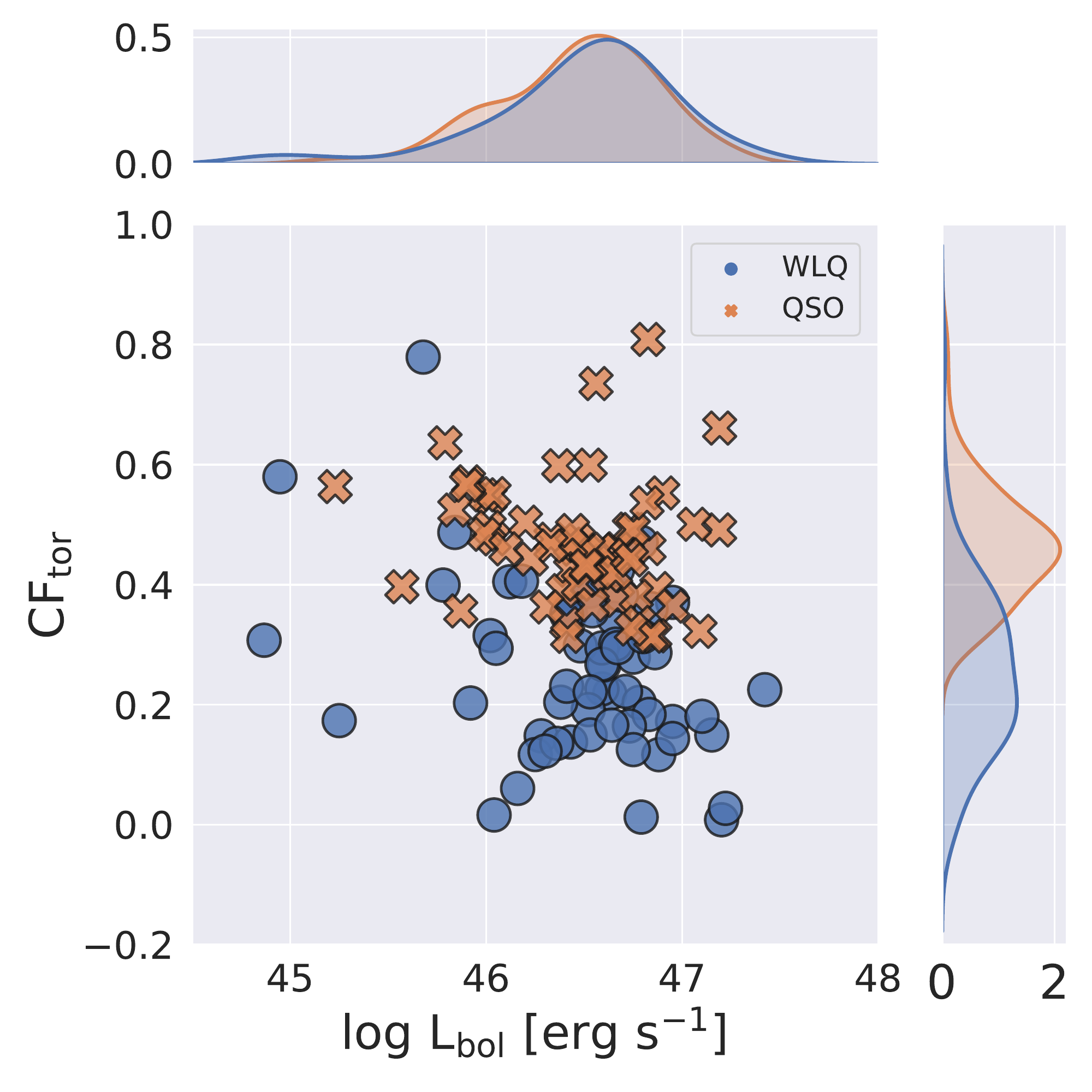}}
   \caption{Correlation between the covering factor of the torus ($CF_{tor}$) and the bolometric luminosity $L_{bol}$. The blue dot represents the WLQs whereas the orange cross represents the control sample of QSOs.}
   \label{fig:6}
\end{figure}
\begin{figure}
   {\includegraphics[width = 9 cm]{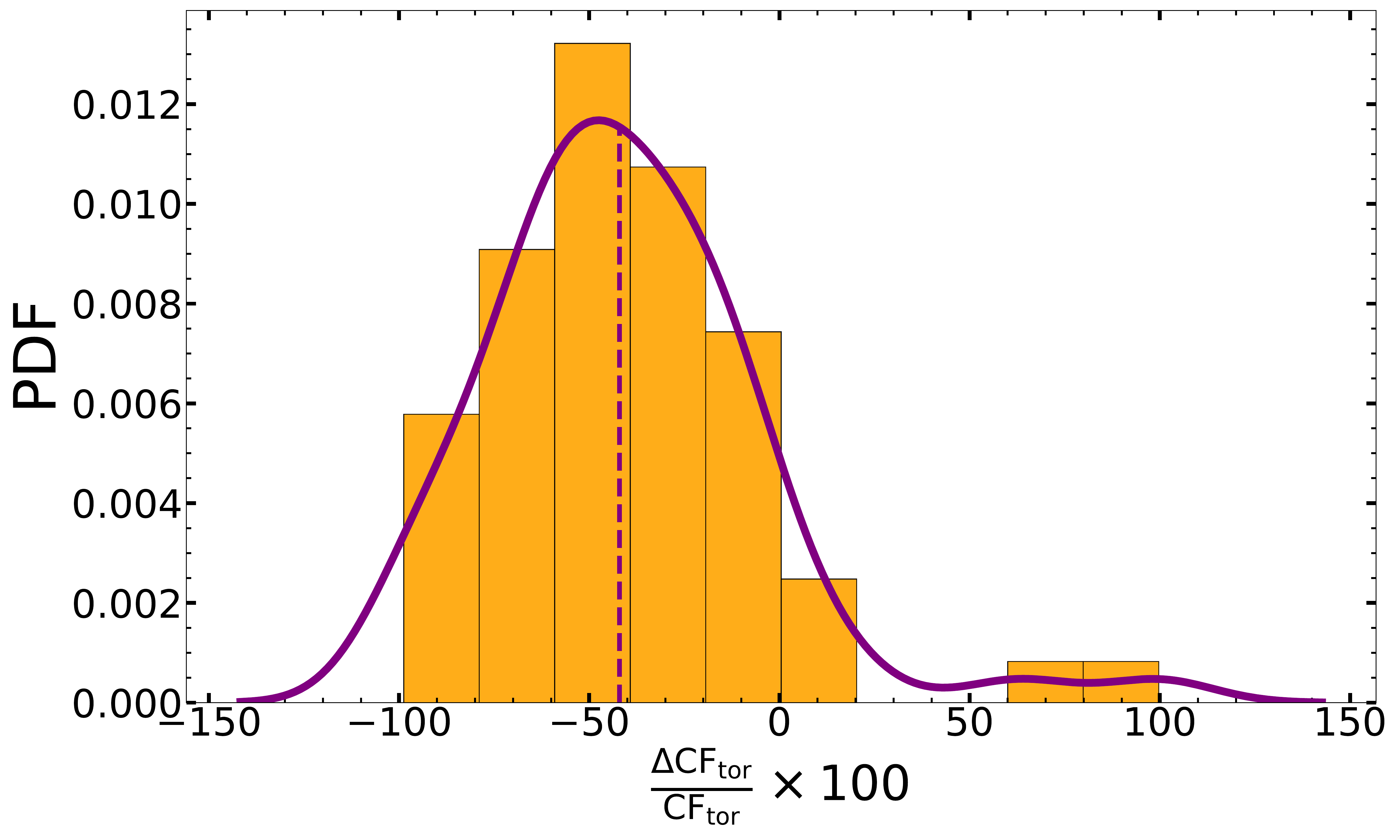}}
   \caption{The kernel smooth probability distribution function (PDF) of the percentage deviation of torus covering factor of WLQ with respect to the QSOs ($\Delta CF_{tor}/CF_{tor}$). The median percentage deviation $\Delta CF_{tor}/CF_{tor}$ is found to be $-42\pm 4\%$.}
   \label{fig:7}
\end{figure}
\section{Analysis and Results}
\label{section:analysis}
In our analysis of SED fitting, we have used the observed flux value of our WLQs sample and control sample derived from their optical SDSS magnitude in u, g, r, i and z bands, and in IR using the WISE magnitude in W1, W2, W3, and W4 bands as listed in Table~\ref{tab:Table_1} and Table~\ref{tab:Table_2}, respectively. Here in the control sample, we have used the median value of the magnitude of the 55 normal QSOs corresponding to each WLQ and the error bar on the median  magnitude is estimated as expected on the mean value by propagating the observed error bar of the individual sources. 
For conversion of SDSS magnitude to flux at the central waveband of the filter, we have used the online platform provided by Gemini\footnote{https://www.gemini.edu/observing/resources/magnitudes-and-fluxes/conversions-between-magnitudes-and-flux} observatory, which is based on SDSS calibrations. For the conversion of WISE magnitude, we have used the formulae $$F_{\nu}[Jy]=F_{\nu 0}\times 10^{(-m_{wise}/2.5)}$$  where $F_{\nu 0}$ is the zero magnitude flux density by assuming the AGN spectrum of the form $F_{\nu } \propto \nu^{-2}$, as detailed in  \citet{wright2010wide}. In Fig.~\ref{fig:fig1} we show the plot of the median values of the observed flux, at the central frequency of the WISE and SDSS filters, both for 69 WLQs as well as their control sample of normal QSOs. As can be seen from this figure, the WISE flux, especially in W1, W2, and W3, is systematically smaller for WLQs in comparison to their control sample of normal QSOs. The quantification of this difference (in observed frame) requires proper decomposition of SED (in rest frame) into its individual emission components, as detailed in the next section. 
\begin{table*}
  \centering
      {
        \setlength{\tabcolsep}{1.2pt}
\caption{{The best fit parameters of SED fit based on {\sc AGNfitter} for our sample of 61 WLQs.}}
\label{tab:Table_3}
\renewcommand{\arraystretch}{1.5}
\renewcommand{\tabcolsep}{2mm}
{\tiny
\begin{tabular}{rlrrrrrccccrrccrr}
\hline
 SN. & Source  Name  &  $\tau$ &     SB &     BB &     GA &     TO &   $EBV_{bbb}$ &   $EBV_{gal}$ &              Age &         N~{\sc h} &   $L_{IR}$ &   $L_{bb}$ &   $L_{bbdered}$ &      $L_{ga}$ &     $L_{tor}$ &      $L_{sb}$ \\
\hline
    1 & J001444.03$-$000018.5 &  8.75 & 2.23 & 2.97 & 3.88 & 2.50  & 0.10  & 0.05 &  5.68 & 21.29 & 46.63 &46.39 &  46.7  & 46.2  &  46.19 & 45.97 \\
2 & J001514.88$-$103043.6 &  7.17 & 1.99 & 2.31 & 4.98 & 2.00& 0.05 & 0.10  &  9.39 & 21.37 & 46.52 &45.64 &  45.8  & 45.07 &  45.45 & 45.84 \\
3 & J001741.87$-$105613.2 &  6.39 & 2.24 & 2.45 & 3.61 & 2.39 &$-$0.04 & 0.27 &  6.47 & 21.38 & 46.27 &46.29 &  46.29 & 45.37 &  46.19 & 45.6  \\
--- & \multicolumn{1}{c}{---} &\multicolumn{1}{c}{---} &\multicolumn{1}{c}{---} &\multicolumn{1}{c}{---} & \multicolumn{1}{c}{---} &\multicolumn{1}{c}{---} &\multicolumn{1}{c}{---} &\multicolumn{1}{c}{---} &\multicolumn{1}{c}{---} &\multicolumn{1}{c}{---} & \multicolumn{1}{c}{---} &\multicolumn{1}{c}{---} &\multicolumn{1}{c}{---} &\multicolumn{1}{c}{---} &\multicolumn{1}{c}{---} &\multicolumn{1}{c}{---}  \\
   \hline
\multicolumn{14}{l}{\textbf{Note:} The entire table is available in online version. Only a portion of this table is shown here to display its form and content.}\\
\end{tabular}
}
}
\end{table*}

\begin{table*}
  \centering
      {
        \setlength{\tabcolsep}{1.2pt}
\caption{{The best fit parameters of SED fit based on {\sc AGNfitter} for the composite spectrum of control sample  corresponding to each of our 61 WLQs.}}
\label{tab:Table_4}
\renewcommand{\arraystretch}{1.5}
\renewcommand{\tabcolsep}{2mm}
{\tiny
\begin{tabular}{llrrrrrrrrrrrrrrr}
\hline
  SN. & Source name$^{*}$  &    $\tau$ &     SB &     BB &     GA &     TO &   $EBV_{bbb}$ &   $EBV_{gal}$ &              Age &         N~{\sc h} &   $L_{IR}$ &   $L_{bb}$ &   $L_{bbdered}$ &      $L_{ga}$ &     $L_{tor}$ &      $L_{sb}$ \\
\hline
    1 & J001444.03$-$000018.5 &  8.02 & 2.25 & 2.92 & 3.97 & 2.8  & 0.1  & 0.1  &  6.48 & 21.76 & 44.01 &46.34 &  46.64 & 46.13 &  46.46 & 43.41 \\
2 & J001514.88$-$103043.6 & 10.16 & 1.67 & 2.28 & 3.58 & 2.28 & 0.05 &$-$0.04 &  8.13 & 21.26 & 43.99 &45.61 &  45.78 & 45.04 &  45.72 & 43.28 \\
3 & J001741.87$-$105613.2 &  6.94 & 2.16 & 2.67 & 3.32 & 2.61 & 0.1  &$-$0.04 &  5.48 & 21.3  & 43.85 &46.21 &  46.52 & 45.97 &  46.42 & 43.22 \\
--- & \multicolumn{1}{c}{---} &\multicolumn{1}{c}{---} &\multicolumn{1}{c}{---} &\multicolumn{1}{c}{---} & \multicolumn{1}{c}{---} &\multicolumn{1}{c}{---} &\multicolumn{1}{c}{---} &\multicolumn{1}{c}{---} &\multicolumn{1}{c}{---} &\multicolumn{1}{c}{---} & \multicolumn{1}{c}{---} &\multicolumn{1}{c}{---} &\multicolumn{1}{c}{---} &\multicolumn{1}{c}{---} &\multicolumn{1}{c}{---} &\multicolumn{1}{c}{---}  \\
\hline
\multicolumn{14}{l}{\textbf{Note:} The entire table is available in online version. Only a portion of this table is shown here to display its form and content.}\\
\end{tabular}
}
}
\end{table*}
{\tiny
\begin{table*}
\caption{The median value of the important parameters of our SED  model fit based on 61 WLQs and using their corresponding composite of their control sample of normal QSOs.} 
\label{tab:parameters}
\renewcommand{\tabcolsep}{6mm}
\begin{tabular}{lrrrrr}
\hline
\textbf{Parameter$^{*}$} & WLQ & \textbf{$\Large{\sigma_{wlq}}$} & QSO &  \textbf{$\Large{\sigma_{qso}}$} & \%deviation \\
  & & & & & $\left[\frac{WLQ-QSO}{QSO}\times 100 \right]$ \\
\hline
\textbf{ SB}                  & 1.99 $\pm$ 0.22      & 1.68  &  1.84 $\pm$ 0.23         & 1.78    & 5.93 $\pm$ 19.17 \\
\textbf{ BB}                  & 2.49 $\pm$ 0.01      & 0.11  &  2.62 $\pm$ 0.001        & 0.003   & $-$2.15 $\pm$ 0.52 \\
\textbf{ GA}                  & 4.29 $\pm$ 0.05      & 0.39  &  3.52 $\pm$ 0.004        & 0.03    & 23.01 $\pm$ 1.44 \\
 \textbf{ TO}                 & 2.35 $\pm$ 0.03      & 0.24  &  2.57 $\pm$ 0.01         & 0.05    & $-$8.59 $\pm$1.20 \\
 \textbf{ EBV\_bb}            & 0.06 $\pm$ 0.01      & 0.08  &  0.10 $\pm$ 0.01         & 0.02    & $-$6.27 $\pm$ 15.47 \\
 \textbf{ EBV\_gal}           & 0.20 $\pm$ 0.01      & 0.09  &  $-$0.034 $\pm$ 0.004    & 0.03    & $-$138.38$\pm$ 130.06  \\
 {\bf  Ldered(0.1$-$1 $\mu m$)}              & 14.09 $\pm$ 0.42     & 3.32  &  19.12 $\pm$ 0.02        & 0.15    & $-$11.81 $\pm$ 1.81 \\
 {\bf  Lga(0.1$-$1 $\mu m$)}  & 5.16 $\pm$ 0.15      & 1.17  &  6.28 $\pm$ 0.01         & 0.06    & $-$5.75 $\pm$7.37 \\
 {\bf  Ltor(1$-$30 $\mu m$)}  & 8.85 $\pm$ 0.51      & 4.002 &  13.99 $\pm$ 0.29        & 2.32    & $-$41.69 $\pm$ 1.59 \\
\hline
\multicolumn{6}{l}{$^{*}${\bf SB, BB, GA, TO} refers to normalization parameters of ``starbusrt", ``big blue bump", ``galaxy" and ``torus" component  respectively.}\\
\multicolumn{6}{l}{$^{*}${\bf EBV\_bb, EBV\_gal} refers to the reddening parameters of ``big blue bump" and ``galaxy" component respectively.}\\
\multicolumn{6}{l}{$^{*}${\bf Ldered, Lga, Ltor} refers to the luminosity of ``deredened big blue bump", ``galaxy" and ``torus" component respectively.}\\
\multicolumn{6}{l}{in unit of $10^{45}$erg/s.}\\
\end{tabular}
\end{table*}
}
\subsection{SED fitting using {\sc AGNfitter}}
The observed value of the flux at the central wavelength of the WISE and SDSS filters is used to fit the SED of each WLQ and the composite of their control sample, consisting of 55 normal QSOs, by using the publically available code {\sc AGNfitter}{\footnote{https://github.com/GabrielaCR/AGNfitter}} as detailed in \citet{calistro2016agnfitter}. In brief, {\sc AGNfitter} disentangles the physical processes responsible for AGN emission, such as the contribution from the stellar populations of the host galaxy, cold dust in star-forming regions, hot dusty torus, and AGN accretion disk. The SED can be constructed using model templates that depict the contribution of each source component at different wavelength ranges, including UV and optical wavelengths to sub-millimeters. \\
The SED of AGNs contains significant features in the ultraviolet to optical region known as the ``Big Blue Bump"(BBB). The BBB is thought to originate from an optically thick accretion disk accreting matter into SMBH, 
whose energy contribution generally peaks at extreme ultraviolet wavelength regime \citep[e.g., see][]{1987ApJ...323..456M}. 
Modeling of BBB in {\sc AGNfitter} is done by using the modified version of the empirical template given by \citet{richards2006spectral} which was derived based on composite spectrum obtained by using 259 Type-1 QSOs. The reddening law for this template was given by \cite{prevot1984typical}, which is found to be effective in treating reddening seen in Type-1 AGNs \citep[e.g., see][]{hopkins2004dust,salvato2008photometric}. 
In their modification, they have independently modeled the mid-IR regime by using the warm dust template \citep[e.g.][]{calistro2016agnfitter}. 
The dusty torus components in {\sc AGNfitter} are modeled using the empirical template given by \citet{silva2004connecting}. For estimating the host galaxy contribution, a stellar population synthesis model of \citet{bruzual2003stellar} is used. Cold dust emission from the star-forming regions is modeled by using  169 templates with a wide range of SED shapes and luminosities \citep[e.g., see][]{chary2001interpreting, dale2002infrared}. 
We have individually fitted the SED for each of the 69 WLQs and their corresponding fit of the median flux of their control sample of normal quasars, using {\sc AGNfitter}, by converting the observed flux value to their rest-frame value. 
Further, to use the ratio of IR-luminosity to the bolometric luminosity ($L_{bol}$) as a measure of covering factor of duty torus, we used the value $L_{bol}$ from the literature. Out of the total 69 sources, the $L_{bol}$ for 61 sources we have taken from \citet[][]{rakshit2020spectral}, and for 5 sources taken from the NASA/IPAC database\footnote{https://ned.ipac.caltech.edu/} and the remaining 3 sources without proper  $L_{bol}$ were excluded from our sample, reducing the sample to 66 sources. Lastly, we noticed in our SED fitted value of torus luminosity for 5 sources does not satisfy the physical condition of $L_{tor}<L_{bol}$, and hence got excluded. This led to the final sample of 61 WLQs sources and the corresponding 61 SED fit of the median flux of the control sample consisting of 55 normal QSOs. The SED fits to our entire sample of 61 WLQs is given in online mode in Fig.~A, however, for illustration in Fig.~\ref{fig:sed}, we have shown our SED comparison of different components for one of the WLQs in our sample, namely {J141141.96+140233.9}, along with the corresponding fit of the median fluxes of its control sample of normal QSOs. 
\subsection{ Dusty torus luminosity and covering factor}
 {\sc AGNfitter} provides a list of various parameters associated with different components. Among them, few important are: (i) contribution from big blue bump (BBB) along with its reddening (E$(B-V)_{bbb}$), (ii) torus emission parameterised by torus luminosity (L$_{tor}(1-30 \mu m)$), torus column density, (iii) stellar emission parametrised by star formation time scale ($\tau[Gyr]$), Galaxy age, reddening ((E$(B-V)_{gal}$)) and luminosity (L$_{gal}(0.1-1 \mu m)$). The best-fit parameters of the SED fit of each of the 61 WLQs in our sample and their corresponding fit of the median fluxes of control sample of normal QSOs are given in Table~\ref{tab:Table_3} and Table~\ref{tab:Table_4}, respectively. These parameters in {\sc AGNfitter} are extracted based on their various realizations and the corresponding associated errors by using 16th and 84th-percentile realizations relative to the median 50th-percentile value, assuming the distribution function to be Gaussian. The scatter plot of $\Delta L_{tor}/L_{tor}(\equiv  [L_W^{tor}-L_Q^{tor}]/L_{Q}^{tor})$  with emission redshift ($z_{emi}$) for our sample of 61 WLQs is shown in Fig.~\ref{fig:3}. As can be seen from this figure that (i) except for a few outliers, all WLQs shows smaller torus luminosity in comparison to their redshift and r-band magnitude matched sample of normal QSOs and (ii) the $\Delta L_{tor}/L_{tor}$  do not show any visual trend with the  $z_{emi}$ with their Pearson correlation coefficient of $-0.094$. This non-significant correlation of $\Delta L_{tor}/L_{tor}$ with $z_{emi}$ allows us to carry out the statistical study of the entire sample even though our WLQs sample belongs to a range in  $z_{emi}$ (0.5 to 3.0).
 The median values of these parameters in 61 WLQs, and the percentage deviation of WLQs w.r.t the SED fit of the median flux of its control sample of normal QSOs are given in Table~\ref{tab:parameters}. In Fig.~\ref{fig:5} we have plotted the kernel smooth probability distribution function (PDF) of $\Delta L_{tor}/L_{tor}$. As can be seen from this figure that $\Delta L_{tor}/L_{tor}$  distribution is clearly showing smaller $L_{tor}$ for WLQs with median value of $\Delta L_{tor}/L_{tor}$ to be $42\pm2\%$. The Gaussian function fit to this distribution results in a typical $1\sigma$ uncertainty of $29\%$. and got the bolometric luminosity from there. In total, we have 66 sources among 69 sources for which we have bolometric luminosity. 
Our estimated $L_{tor}$ can also be used to estimate the covering factor of a torus ($CF_{tor}$), in conjunction with the available $L_{bol}$ measurements, as $CF_{tor} =L_{tor}/L_{bol}$ \citep[e.g., see ][]{zhang2016covering}. The plot of $CF_{tor}$ with $L_{bol}$ along with distributions of $CF_{tor}$ both for WLQs as well as normal QSOs are shown in  Fig.~\ref{fig:6}. As can be seen from this figure, the $L_{bol}$ distribution of WLQs and QSOs are similar with KS-test $P_{null}$ of $0.82$, but the distribution of $CF_{tor}$ is significantly different with KS-test $P_{null}$ of $4.27\times 10^{-14}$, being systematically lower for WLQs. Since $L_{bol}$ also includes the contribution from the $L_{tor}$, to check its impact on the derived result we also estimated the covering factor as $L_{tor}/(L_{bol}$-$L_{tor})$ and found the percentage decrement of 61\% which is higher than the 42\% based on the use of $L_{tor}/L_{bol}$. However, for the sake of comparison with earlier studies, we have adhered to the definition of covering factor of dusty torus as $L_{tor}/L_{bol}$. 
In Fig.~\ref{fig:7} we plot the kernel smooth probability distribution function  of the percentage deviation of torus covering factor, $\Delta CF_{tor}/CF_{tor}(\equiv [CF_W^{tor}$-$CF_Q^{tor}]/CF_{Q}^{tor})$, which clearly shows the decrement in the covering factor of WLQs with a median value of about $-42\pm 4\%$.
\section{DISCUSSION AND CONCLUSIONS}
\label{section:discussion}
  With the advent of large spectroscopic surveys, various techniques have been used to unravel the nature of the enigmatic population of WLQs, especially in the context of testing two main scenarios: (i) insufficient ionizing photons as the cause of weak emission lines in WLQs and (ii) WLQs being in the early phase of quasar evolution where BLR is yet to be fully developed.
 As suggested by \citet{gaskell2009broad}, such under-developed BLR will also suggest the low covering factor of the dusty torus, which will result in the decrement of IR-emission (e.g., see Sec.~\ref{Sec 1}). To test this hypothesis, we have carried out the SED fitting of 61 WLQs and compared them with the control sample of the normal QSOs, matching in redshift and r-band magnitude (e.g., see Sec.~ \ref{section:sample}). With our detailed SED modeling (e.g., see Sec.~\ref{section:analysis}) the median value of $\Delta L_{tor}/L_{tor}$ is found to be $-0.42\pm 0.02$ with typical r.m.s scatter from the median of about 0.29 (e.g., see Fig.~\ref{fig:5}), suggesting that the torus luminosity of WLQs is about 42\% lower in comparison to the redshift and (optical) luminosity matched sample of the normal QSOs.\par
Our results are consistent with many earlier studies suggesting that WLQs are in the AGN stage of the early phase of their evolution, where BLR has not yet fully developed and hence results in weak emission lines. For instance, \cite{DiamondStanic2009ApJ...699..782D} reported that two WLQs are fainter in the IR band by 30-40\% (e.g., see Sec.~\ref{Sec 1}). 
On the other hand, \citet{zhang2016covering} have also carried out SED fitting of 73 WLQs, where although they did not find any significant difference in the SED of the WLQs and normal QSOs, their results were also supporting the evolution scenario for WLQs, where due to much larger scale of torus it should form before the formation of BLR. 
It may be noted that the procedure of SED fitting used in \citet{zhang2016covering}, only involved components of power-law and black-body. However, in our SED fitting, we have carefully decomposed the various components of emission, which has allowed us to separate out the emission coming from the dusty torus. For comparison with normal QSOs, our study for the first time makes use of control sample matching in redshift and r-band magnitude. These two improvements have perhaps allowed us to detect the difference in torus luminosity of WLQs and normal QSOs, which was missing in the oversimplified aforementioned model used in \citet{zhang2016covering}, in spite of the fact that the majority of sources in our sample overlap with their sample. Moreover, we also stress here that our results presented here are based on the comparison of WLQs SED fit with their redshifted and r-band luminosity matched control sample. As a result, any unknown systematics in observation and/or in SED modeling will not affect our above results of percentage deviation both regarding the torus (IR) luminosity as well as in the torus covering factor. \par
We also note that the emission redshift of our WLQs sample (hence of the redshift-matched control sample as well) varies over a wide range from 0.5 to 3.0. However, as can be seen from Fig.~\ref{fig:3}, the relative change of torus luminosity does not show any correlation with redshift. This non-evolution of $L_{tor}$ with redshift enables us to compare statistically the distribution of the entire sample of WLQs with the control sample of normal QSOs, resulting in the aforementioned 42$\pm$2\% IR-luminosity decrement in WLQs w.r.t the control sample of normal QSOs, comprising a 3$\sigma$ range of 36\% to 48\%. For individual sources also, it can be noted from Fig.~\ref{fig:3} that the $L_{tor}$ of WLQs is found to be consistently smaller in comparison to the control sample of normal QSOs, for all the individual sources in our sample except few outliers (being less than 10 percent). The distribution of $\Delta L_{tor}/L_{tor}$ is well fitted with Gaussian with a best fit mean of $-0.39\pm 0.04$ and $\sigma$ of 0.29. The large value of $\sigma$ ($= 0.29$) suggests that individual measurements of  $\Delta L_{tor}/L_{tor}$ still have high uncertainty, which needs to be reduced either statistically by enlarging the sample size as employed in our study here or by using a large number of photometric data points in SED fit, such as by adding UV and X-ray data in future as well. It can also be noted that the error on the median (i.e., $\sim 2\%$) obtained after error propagation is almost half of the value obtained on the mean (i.e., $\sim 4\%$) based on the $\sigma$ of the  $\Delta L_{tor}/L_{tor}$ distribution. This could be either due to the error under-estimation in flux (i.e., magnitude) or over-estimation of the r.m.s scatter due to a few outliers. Nevertheless, to be on the conservative side, even if we took the higher conservative uncertainty (i.e., 4\% on median value), the decrement detected here in IR-luminosity in WLQs w.r.t the control sample of normal QSOs still deviates from zero at high significance with 3$\sigma$ range of $-$30\% to $-$54\% around the measured value of $42\%$.\par 
Based on the measured width of the emission line, we notice that all the sources in our sample are Type-I AGNs, so it is unlikely that the decrement in IR flux in WLQs can be attributed to any an-isotropic emission of BBB component if it arises from the accretion disk, as such an effect would be dominant in the sample dominated by Type-II AGNs. We also note that  the photometric data points used in optical-based on SDSS and in IR-based on WISE are not simultaneous. As a result, our SED fit implicitly assumes that the variation of WLQs and/or of QSOs used in our sample is not very significant. This could be a reasonable assumption as both WLQs and normal QSOs show nominal variation, unlike the highly variable sources such as Bl Lacs and blazar \citep[e.g., see][]{DiamondStanic2009ApJ...699..782D, shemmer2010weak,Nikolajuk2012MNRAS.420.2518N,2018MNRAS.479.5075K}. Moreover, our study used a reasonably large sample, so any nominal variation between the epoch of IR and optical photometric observation in any individual source should not have any significant effect on our above statistical result. 
Further, we also noted that in our r-band magnitude matched sample of WLQ and normal QSOs, the possibility of the emission line flux contribution will be more in QSOs (due to strong emission lines) than in WLQs (due to weak emission lines). 
This implies that the intrinsic continuum of WLQs (having less contamination of emission lines) may be brighter (in optical and hence in IR as well) than their control sample of normal QSOs (matched in r-band magnitude). As a result, the actual decrement of IR flux in WLQs might be more than 42\% as we have found here. The other possibility, in the context of our result of WLQs being in the early phase of AGN evolution, maybe the episodic phase of the AGN. However, observationally, AGN’s episodic activities are mainly inferred based on episodic changes in their brightness or recurring jet activities \citep[e.g., see][]{saikia2010recurrent,khrykin2016he}, typically giving an episodic time scale of a few million years. In this scenario, if the WLQs phase is a new emerging episode of AGN activity, then it may be in the phase of growing brightness due to restarted AGN activities without destroying the BLR. This might lead to more decrement in continuum and hence increase the measured equivalent width of the emission line, contrary to the systematically smaller EW in WLQs. It suggests that the WLQs maybe perhaps in the early phase of AGN’s formation rather than belonging to any episodic phase.\par 
The lower $L_{tor}$ will also have additional consequences in lowering the torus covering factor ($CF_{tor}$) estimated as $L_{tor}/L_{bol}$. As can be seen from Fig.$~$\ref{fig:6} that the distribution $CF_{tor}$ for WLQs is significantly lower than that of the normal QSO, with KS-test $P_{null}$ of $4.27\times10^{-14}$. This results in a median percentage decrement of $\Delta CF_{tor}/CF_{tor}= 42\pm 4 \%$ (e.g., see Fig.$~$\ref{fig:7}) with a typical r.m.s scatter from the median of about $35\%$, suggesting that the torus covering factor of WLQs on an average is about $42\%$ smaller with a $3\sigma$   of $-30\%$ to $-54\%$ than that of the normal QSOs of similar redshift and optical luminosity. The covering factor of the torus and the BLR has to be of a similar order \citep[e.g.,][]{1989ApJ...342...64A, 1993ApJ...404L..51N,2006ApJ...639...46S,gaskell2007ngc,gaskell2009broad}, so from our above results, we conclude that the BLR in the WLQs is underdeveloped and it can be a dominant cause of the weakness of their emission line. This gives support to the model of WLQs proposed based on their evolution scenario where WLQs are a special stage in the early phase of AGNs. 
\section*{Acknowledgments}
\label{section:acknowledge}
We thank the referee Professor Robert R. J. Antonucci for his critical comments and helpful suggestions on the manuscripts. RK and HC  are grateful to Gabriela Calistro-Rivera for making the code {\sc AGNfitter} public, and  IUCAA for the hospitality under IUCAA associate programme.\\
\section*{DATA AVAILABILITY}
The data used in this study are publicly available in the SDSS  DR14 and WISE All-Sky Data Release.
\bibliography{references}
\label{lastpage}
\end{document}